\documentstyle[epsf]{mnsty}

\author [Hoekstra et al.]
{H.~Hoekstra$^1$, T.S. van Albada$^1$, \& R. Sancisi$^{1,2}$\\
        $^1$~Kapteyn Astronomical Institute, University of Groningen, 
        Postbus 800, 9700 AV Groningen, The Netherlands \\
        $^2$~Osservatorio Astronomico
	Via Ranzani 1, I-40127 Bologna, Italy}

\title{On the apparent coupling of neutral hydrogen and dark matter in spiral 
galaxies}

\begin{document}

\maketitle

\begin{abstract}

We have studied a mass model for spiral galaxies in which the dark
matter surface density is a scaled version of the observed HI surface
density.  Applying this mass model to a sample of 24 spiral galaxies
with reliable rotation curves one obtains good fits for most galaxies.
The scaling factors cluster around 7, after correction for the presence
of primordial helium. But for several cases different, often larger,
values are found. For galaxies that can not be fitted well the
discrepancy occurs at large radii and results from a fairly rapid
decline of the HI surface density in the outermost regions.  Because
of such imperfections and in view of possible selection effects it is
not possible to conclude here that there is a real coupling between HI
and dark matter in spiral galaxies.

\end{abstract}

\section{Introduction}

The first indications that spiral galaxies are surrounded by unseen
matter date to the early seventies (Freeman 1970; Roberts 1975;
Shostak 1973). Later 21-cm line studies, especially those of Bosma
(1978,1981) and Begeman (1987), have all confirmed this
discovery. Dark matter manifests itself already in the outer(most)
parts of the luminous body of spiral galaxies (Rubin 1978; Kent 1986),
but its main signature is the general flatness of rotation curves well
outside the optical radius.

Bosma (1978, 1981) noted a curious result for the galaxies in his sample:
the total surface density of matter needed to explain the observed
rotation curve in the outer region, i.e. outside the optical radius,
was roughly proportional to the surface density of neutral hydrogen.
Later work has shown that this proportionality not only holds for the
high surface brightness (HSB) galaxies studied by Bosma, but also for
dwarf galaxies (Carignan 1985a; Carignan \& Puche 1990; Jobin \&
Carignan 1990).

These results, and other considerations, have led Pfenniger et
al. (1994a, b) to the hypothesis that the dark matter surrounding
spiral galaxies consists of cold gas, mainly in the form of molecular
hydrogen. The spatial distribution of this cold gas should be similar
to that of the observed neutral hydrogen.

As the contribution of luminous matter to the total mass density is
negligible in the outer parts (under the assumption that the
mass-to-light ratio of the luminous matter is independent of radius),
Bosma's relationship is equivalent to a proportionality of dark matter
and neutral hydrogen.

This apparent coupling of dark matter and neutral hydrogen deserves
closer scrutiny, because it seems at odds with the current view that
dark haloes are triaxial (but not extremely flattened) and that they
consist of non-baryonic material. Rather, this coupling suggests
possible support for the hypothesis put forth by Pfenniger et
al. (1994a).

Several questions should be considered: (i) How tight is this relation
between the surface densities of HI and dark matter for the galaxies
with the best available data?  (ii) To what extent does the relation
depend on the use of the so-called maximum disc assumption? (iii) Does
the relation also hold for the low surface brightness (LSB) galaxies?

In this paper we investigate the apparent coupling of dark matter and
neutral hydrogen for a sample of 24 spiral galaxies with good rotation
curves. The sample contains both HSB and dwarf galaxies.
Unfortunately the quality of the rotation curves for LSB galaxies is
still rather poor and we have decided not to include those galaxies in
the present discussion.

We wish to stress at the outset that there is a potentially powerful
selection effect that may cause a relationship between the surface
densities of HI and dark matter for the galaxies in our sample. This
is because the HI surface density distributions of the galaxies in our
sample have the common characteristic that the highest values in the
inner regions, as well as the lowest values in the outer regions are
similar from galaxy to galaxy (the latter being linked to the sensitivity
reached in the observations). As a result, a large radial extent
implies a large radial scale length of HI and vice versa. 
In section~5 we discuss how this can lead to a proportionality
between dark matter and HI.

\section{Mass model}

To study the mass distribution, and in particular the apparent
coupling between the dark matter and HI, we use axisymmetric
distributions of matter containing three components. These
are used to model the derived rotation curves. The various components
of the mass model are:

\begin{enumerate}
\item[{\it Stars:}] {The stars are distributed in a disc. If present,
we also include a bulge component. The radial surface density
profiles of the stellar components are estimated using luminosity
profiles from the literature, assuming a mass-to-light ratio
independent of radius. We further assume that the stellar disc has an
exponential vertical density profile, with a scale height that is one
fifth of the optical disc scale length (van der Kruit \& Searle
1981a,1981b).}

\item[{\it Gas:}] { We use the observed radial surface density
distribution of HI and we assume that the vertical distribution mimics
that of the stars. In reality most of the neutral hydrogen lies in a
thin disc, but our results are not sensitive to the detailed
$z-$distribution because in the inner regions (where the thick disc
description for HI does not hold) the contribution of HI to the
gravitational field can be safely neglected. Apart from neutral
hydrogen, detectable through the 21-cm line, galaxies contain molecular
gas. We assume that the bulk of this molecular gas follows the surface
density profile of the stellar component. In this way the presence of
molecular gas is taken into account in the value of the mass-to-light
ratio of the stellar component.}

\item[{\it Dark matter:}] {Usually a spherical halo is used to
describe the density distribution of the dark matter. If the dark
matter and neutral hydrogen are indeed coupled, as Pfenniger et
al. (1994a) argue, it is reasonable to assume that their spatial
distributions are similar. In this paper we therefore assume that the
surface density of dark matter is proportional to the observed surface
density of neutral hydrogen. The vertical scale height of the dark
matter, like that of HI, is taken equal to that of the stars.  For the
discussion in this paper the choice of the scale height is fairly
arbitrary because the changes in calculated rotation velocities, given
a surface density profile, are small when the scale height is varied.
But the derived total masses do depend on scale height.}
\end{enumerate}

The mass model can be fitted to the observations in two ways: one can
fit the model rotation curve to the observed one, or one can compare
the total surface density, calculated from the rotation
curve, to the measured surface densities of stars and HI.  
We prefer to fit the circular velocities calculated from the mass model
to the observed rotation curves. To do so, we compute the rotation
velocity of a disc using (Cuddeford 1993):

\begin{equation}
V_c^2(R)=2\pi G R \int \limits_{0}^{\infty} dk {S(k)\over{1+k\epsilon}}
k J_{1}(kR),
\end{equation}

\noindent where $J_\nu(x)$ is the Bessel function of order $\nu$ and where 
$S(k)$ is given by:

\begin{equation}
S(k)=\int \limits_{0}^{\infty} dR' \Sigma (R')R'J_{0}(kR').
\end{equation}

\noindent In these equations $\Sigma(R)$ is the surface density of the
disc as a function of radius $R$, and $\epsilon$ is the scale height
of the disc, assuming a density profile given by
$\rho(x,y,z)=\Sigma(R) {\rm e}^{-z/|\epsilon|}$.

Equation (1) requires that the surface density is known out to
infinity. This can only be achieved through extrapolation, for which
we assumed an exponential law.  At large radii the surface densities
are low, thus giving rise to minor contributions to the
integral. Finally, the model rotation curve is calculated by adding
the circular velocities, calculated for the different components, in
quadrature:

\begin{footnotesize}
\begin{equation}
V_{\rm tot}^2=\left({\frac{\Sigma_{\rm dark}}{\Sigma_{\rm HI}}+1}\right) 
V_{\rm HI}^2+ \left({\frac{\rm M}{\rm L_B}}\right)_{\rm b} V_{\rm bulge}^2+
\left({\frac{\rm M}{\rm L_B}}\right)_{\rm d}
 V_{\rm disc}^2,
\end{equation}
\end{footnotesize}

\noindent where $\Sigma_{\rm dark}$ denotes the surface density of the dark 
matter component and where $\Sigma_{\rm HI}$ is the surface density of the HI 
disc.

In making the fits we have three free parameters at our disposal:
$\Sigma_{\rm dark}/\Sigma_{\rm HI}$, $({\rm M/L_B})_{\rm bulge}$ and
$({\rm M/L_B})_{\rm disc}$. For late type galaxies, that show no clear
signs of a bulge, we omitted the (spherical) bulge contribution. The
fits were done by eye, in such a way that the model rotation curve
explained the observed rotation curve as far out as possible.

\begin{table*}
{\footnotesize
\centering
\caption{Basic properties of the selected galaxies}
\smallskip
\begin{tabular}{llrrrl||llrrrl}
\hline
Name & Type & Distance & $M^{b,i}_B$ & $V_{\rm max}$ & Ref. & Name & Type & 
Distance & $M^{b,i}_B$ & $V_{\rm max}$ & Ref. \\
(1)  & (2)  & (3)      & (4)         & (5)       & (6)     & (1)  & (2)  & (3)      
& (4)         & (5)       & (6)     \\
\hline
     &      & [Mpc]    & [mag]       & [km/s]    &         &      &      & [Mpc]    
& [mag]       & [km/s]    &         \\
\hline
DDO 154  & Irr & 4.0  & --13.8 & 48  & 1, 2   & NGC 3109 & Sm  & 1.7  & --16.8 & 
67  & 14, 15 \\
DDO 168  & Irr & 3.5  & --15.2 & 55  & 3      & NGC 3198 & Sc  & 9.4  & --19.4 & 
157 & 14, 16 \\
DDO 170  & Sm  & 12.0 & --14.5 & 66  & 4      & NGC 5033 & Sbc & 11.9 & --20.2 & 
225 & 10, 13 \\
NGC 55   & Sm  & 1.6  & --18.6 & 87  & 5      & NGC 5371 & Sb  & 34.8 & --21.7 & 
242 & 13, 20 \\
NGC 247  & Sd  & 2.5  & --18.0 & 108 & 6, 7   & NGC 5533 & Sab & 55.8 & --21.4 & 
302 & 3, 9, 18 \\
NGC 300  & Sd  & 1.8  & --17.8 & 97  & 8, 7   & NGC 5585 & Sd  & 6.2  & --17.5 & 
92  & 19     \\
NGC 801  & Sc  & 79.2 & --21.7 & 238 & 3, 10  & NGC 6503 & Scd & 5.9  & --18.7 & 
121 & 12, 20 \\
NGC 1560 & Sd  & 3.0  & --15.9 & 79  & 12     & NGC 6674 & Sb  & 49.3 & --21.6 & 
291 & 3,9    \\
NGC 2403 & Scd & 3.3  & --19.3 & 136 & 13, 14, 23 & NGC 6946 & Scd & 10.1 & --21.4 & 
170 & 11, 17 \\
NGC 2841 & Sb  & 18.0 & --21.7 & 326 & 13, 14 & NGC 7331 & Sb  & 14.9 & --21.4 & 
257 & 12, 14 \\
NGC 2903 & Sbc & 6.4  & --20.0 & 216 & 13, 14 & UGC 2259 & Sdm & 9.8  & --17.0 & 
90  & 21, 14 \\
NGC 2998 & Sc  & 67.4 & --21.9 & 214 & 3, 10  & UGC 2285 & Sc  & 78.7 & --22.8 & 
298 & 10, 22 \\
\hline
\end{tabular}
\parbox{16cm}{ Galaxy types, listed in column (2), have been taken
from Broeils (1992a) or from the original papers. Distances as given in
Broeils (1992a) have been adopted ($H_o={\rm 75 km/s/Mpc}$).  Total
blue magnitudes have been taken from Broeils (1992a) as well, when
available. If not, we used values listed in the original
papers. $V_{\rm max}$ is the maximum observed rotation velocity of the
galaxy.}
\begin{tabbing} nr\hspace{0.5cm} \= nr\hspace{5cm}\qquad \= nr\hspace{4cm} \= 
nr\hspace{5cm} \= \kill
\> {\it Key~to~the~references:}		\> 				\> 			\\
\> 1 Carignan \& Freeman (1988) 	\>  9 Broeils \& Knapen (1991) 	\> 17 Kamphuis (1993) \\
\> 2 Carignan \& Beaulieu (1989) 	\> 10 Kent (1986) 		\> 18 Kent (1984) \\
\> 3 Broeils (1992a) 			\> 11 Carignan et al. (1990) 	\> 19 C\^{o}te, Carignan \& Sancisi (1991) \\
\> 4 Lake, Schommer \& van Gorkom (1990)\> 12 Broeils (1992b) 		\> 20 Wevers (1984) \\
\> 5 Puche, Carignan \& Wainscoat (1991)\> 13 Begeman (1987) 		\> 21 Carignan, Sancisi \& van Albada (1988) \\
\> 6 Carignan \& Puche (1990) 		\> 14 Kent (1987) 		\> 22 Roelfsema \& Allen (1985) \\
\> 7 Carignan (1985b) 			\> 15 Jobin \& Carignan (1990)  \> 23 Sicking (1997) \\
\> 8 Puche, Carignan \& Bosma (1990) 	\> 16 Begeman (1989)   		\>                               
\\
\end{tabbing}
}
\end{table*}

\section{The sample}

The sample used for this study is essentially that of Broeils (1992a), with
galaxies characterized by the following:

\begin{itemize}
\item{The HI rotation curve has been measured with the Westerbork
Synthesis Radio Telescope or with the Very Large Array.}
\item{The galaxy has a smooth and fairly symmetric velocity field and gas
distribution.}
\item{High precision photometric data are available.}
\item{Inclination and position angle have been derived kinematically,
using a tilted ring analysis of the velocity field.}
\end{itemize}

We have included NGC~5371 (Begeman 1987), which was not used by
Broeils `because of its clumpy gas distribution and asymmetric
velocity field'. For our purpose the properties of this galaxy are
sufficiently regular. We have further added NGC~6946 using data from
Kamphuis (1993). NGC~1003 from Broeils' sample could not be used,
because the data are not readily available.  The selected galaxies are
listed in Table~1, together with some of their basic properties. The
distances given in this table are based on the
values\footnote{Adopting $H_o =$ 75 km/s/Mpc.} given in Broeils
(1992a). For NGC~5371 and NGC~6946 distances from Begeman (1987) and
Carignan et al. (1990) have been used.

The galaxies in Table~1 span a considerable range in Hubble type, from
Sab to Irr. Although most galaxies are of type Sc or later, the sample
allows for a systematic study of galaxy properties from types Sb to
Irr. Note that our sample does not include LSB galaxies.

\begin{figure*}
\begin{center}
\leavevmode
\hbox{%
\epsfxsize=17.5cm
\epsffile[40 170 590 690]{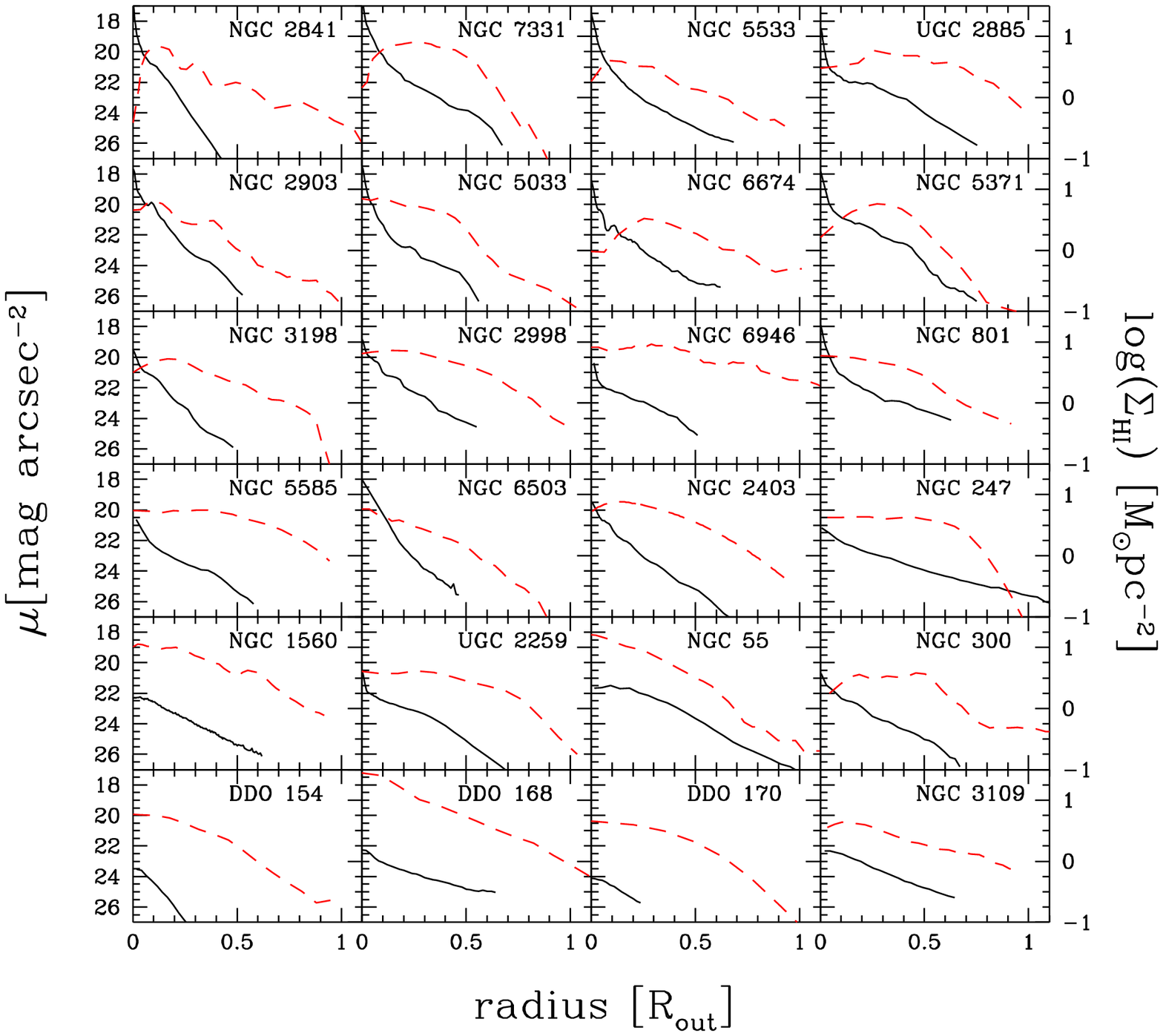}}
\end{center}
\caption{Mosaic of $r$-band surface brightness profiles (solid lines) and
HI surface density profiles (dashed lines) of the galaxies in our sample.
The profiles are plotted against radius expressed in $R_{\rm out}$, the 
outermost point of the rotation curve}\label{profHI}
\end{figure*}

The light profiles and HI surface density distributions for the
galaxies in Table~1 are shown in Figure~\ref{profHI}. Apart from the
occasional presence of a bulge, most light profiles are essentially
exponential. The HI surface density distributions show a larger
variety of shapes, but the maximum surface densities reached in the
inner regions are remarkably similar: about $6~{\rm M}_\odot/{\rm
pc}^2$. Note the larger radial extent of the neutral hydrogen
component compared to the light, and the steepening of the HI profile
in the outer region for several galaxies. However, for some of the
dwarf galaxies the scalelengths of light and HI gas are quite similar.

\begin{figure*}
\begin{center}
\leavevmode
\hbox{%
\epsfxsize=17.5cm
\epsffile[40 170 590 690]{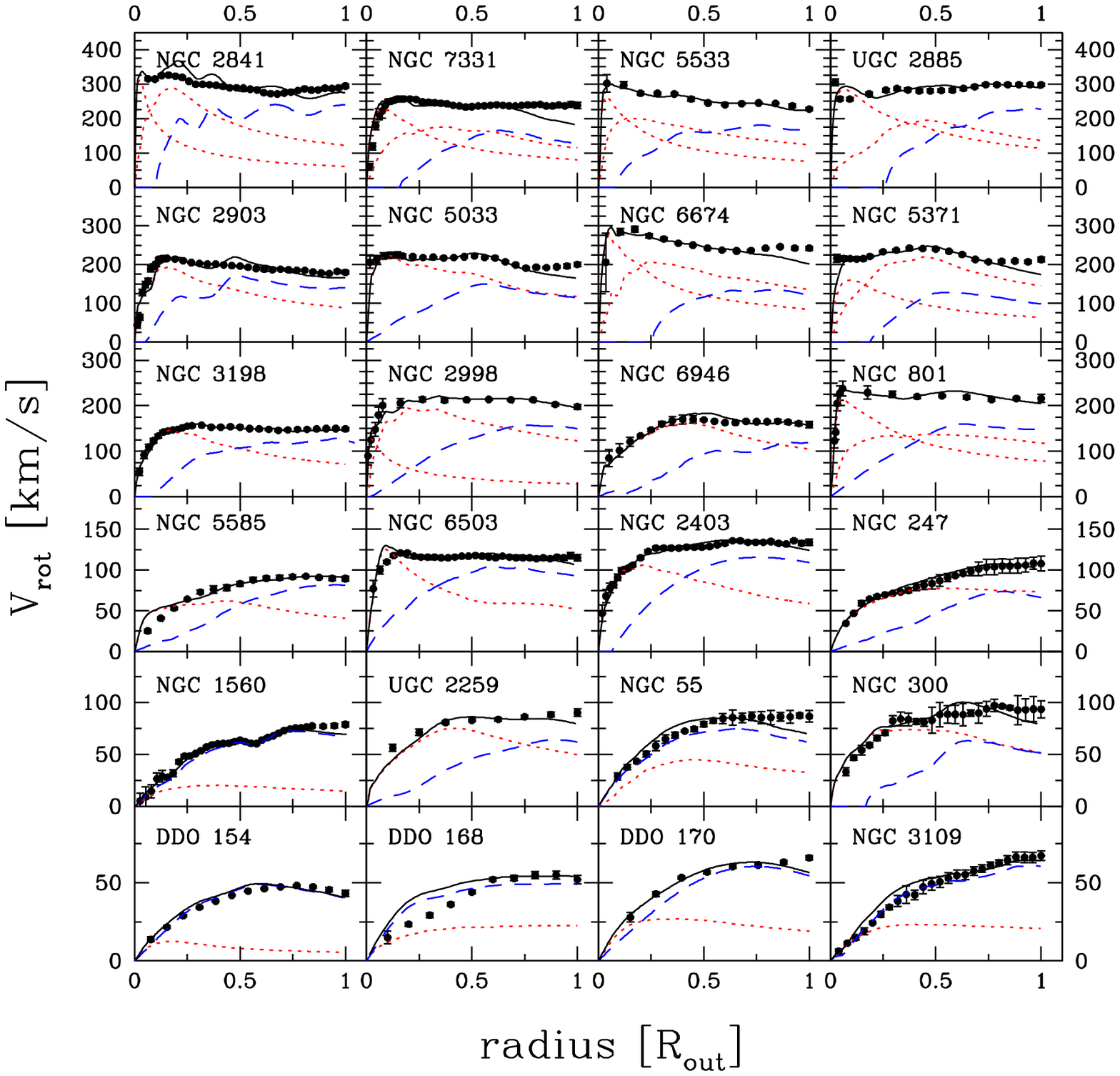}}
\end{center}
\caption{Mosaic of rotation curves. The published rotation curves are
the dots with error bars. The model fits to these rotation curves are
indicated by the solid lines. The dotted lines denote the disc and, if
included, the bulge contribution.  The dashed lines correspond to the
scaled HI contribution. The curves are plotted against radius
expressed in $R_{\rm out}$, the outermost point of the rotation
curve. Error bars in the published rotation curves are usually formal
errors from the tilted ring model fits.  The true uncertainties are
probably somewhat larger, perhaps a factor of 2, than the quoted
errors.}\label{rotcur}
\end{figure*}

The HI surface density profiles in Figure~\ref{profHI} have been
derived from 21-cm line synthesis observations. For large galaxies the
HI flux in the outer parts may therefore have been underestimated.
As a consequence of the missing ``large scale length'' HI components,
rotation curves in Figure~\ref{rotcur} may show a decrease in the outermost
parts. Comparison of the total flux with that measured with
single dish telescopes (Rots 1980) shows that the profiles of NGC~55
and~300 may well suffer from this. This may also be true for NGC~3109.

\section{Results of mass model fits}

In this section we present and discuss the fits of the 
adopted mass model to the rotation curves of the 24 galaxies in
Table~1. In Table~2 we list the values of the fitted
parameters and the inferred masses of the various components.

\begin{table*}
{\footnotesize
\caption{Results of mass model fits}
\smallskip
\begin{center}
\begin{tabular}{lcccccccc}
\hline
\multicolumn{1}{c}{Name} & 
\multicolumn{1}{c}{Type} & 
\multicolumn{1}{c}{${\rm (M/L_B)_{disc}}$} & 
\multicolumn{1}{c}{${\rm (M/L_B)_{bulge}}$} & 
\multicolumn{1}{c}{$\Sigma_{\rm dark}/\Sigma_{\rm HI}$} & 
\multicolumn{1}{c}{${\rm M}_{\rm disc}$} &
\multicolumn{1}{c}{${\rm M}_{\rm bulge}$} & 
\multicolumn{1}{c}{${\rm M}_{\rm HI}$} & 
\multicolumn{1}{c}{${\rm M}_{\rm tot}$} \\
\multicolumn{1}{c}{(1)} & 
\multicolumn{1}{c}{(2)} & 
\multicolumn{1}{c}{(3)} & 
\multicolumn{1}{c}{(4)} &
\multicolumn{1}{c}{(5)} & 
\multicolumn{1}{c}{(6)} &
\multicolumn{1}{c}{(7)} &
\multicolumn{1}{c}{(8)} &
\multicolumn{1}{c}{(9)} \\
\hline
DDO 154  & Irr & 1.0 & -   & 8.0  & 0.052 & -   & 0.28 & 2.6 \\
DDO 168  & Irr & 1.2 & -   & 7.0  & 0.23  & -   & 0.21 & 1.9 \\
DDO 170  & Sm  & 3.5 & -   & 9.5  & 0.34  & -   & 0.45 & 5.3 \\
NGC 55   & Sm  & 0.5 & -   & 7.0  & 2.2   & -   & 0.9  & 9.4 \\
NGC 247  & Sd  & 4.4 & -   & 8.0  & 11    & -   & 0.8  & 18 \\
NGC 300  & Sd  & 2.8 & -   & 11.0 & 5.9   & -   & 0.7  & 14 \\
NGC 801  & Sc  & 2.8 & 4.5 & 9.0  & 160   & 72  & 22   & 450 \\
NGC 1560 & Sd  & 1.0 & -   & 7.0  & 0.36  & -   & 0.82 & 6.9 \\
NGC 2403 & Scd & 1.8 & -   & 9.5  & 15    & -   & 3.1  & 46 \\
NGC 2841 & Sb  & 5.3 & 3.1 & 23.0 & 280   & 68  & 16   & 730 \\
NGC 2903 & Sbc & 2.7 & -   & 37.0 & 43    & -   & 2.4  & 130 \\
NGC 2998 & Sc  & 1.5 & 5.0 & 7.0  & 130   & 8.5 & 23   & 320 \\
NGC 3109 & Sm  & 0.8 & -   & 13.5 & 0.66  & -   & 0.49 & 7.8 \\
NGC 3198 & Sc  & 3.9 & -   & 12.5 & 35    & -   & 5.0  & 100 \\
NGC 5033 & Sbc & 5.8 & -   & 12.0 & 110   & -   & 6.5  & 190 \\
NGC 5371 & Sb  & 2.9 & 3.3 & 8.0  & 120   & 60  & 8.6  & 260 \\
NGC 5533 & Sab & 5.0 & 8.0 & 16.5 & 220   & 96  & 25   & 750 \\
NGC 5585 & Sd  & 2.0 & -   & 9.0  & 3.2   & -   & 1.3  & 16 \\
NGC 6503 & Scd & 1.2 & -   & 17.5 & 5.6   & -   & 1.6  & 35 \\
NGC 6674 & Sb  & 4.5 & 7.0 & 9.0  & 210   & 110 & 30   & 620 \\   
NGC 6946 & Scd & 1.2 & -   & 4.5  & 53    & -   & 20   & 160 \\
NGC 7331 & Sb  & 4.6 & 1.8 & 8.0  & 110   & 59  & 11   & 270 \\
UGC 2259 & Sdm & 4.0 & -   & 10.0 & 3.9   & -   & 0.42 & 8.5 \\
UGC 2885 & Sc  & 1.5 & 7.5 & 9.5  & 270   & 210 & 44   & 940 \\
\hline
\end{tabular}
\parbox{11.7cm}{\noindent Mass-to-light ratios in columns (3) and (4)
are given in units of ${\rm M}_\odot / {\rm L_{B_\odot}}$; masses in
columns (6), (7), (8) and (9) are given in $10^9 {\rm
M}_\odot$. Column (9) gives the total mass of the model.}
\end{center}
}
\end{table*}

The models are compared to the observed rotation curves in
Figure~\ref{rotcur}. The radius is expressed in units of $R_{\rm
out}$, the radius of the last measured point of the rotation
curve. The circular velocities for the stellar component(s) (dotted)
and the scaled HI component (dashed) are also plotted.

We see that most of the 24 rotation curves can be fitted rather well
over their full extent by scaling up the HI surface density. The
poorer fits can be categorized as follows:

\begin{enumerate}

\item[(1)]{The model curve does not agree with the observed rotation
curve in the inner region. This can be due to an incorrect bulge
contribution. 

For dwarf galaxies the inner parts of the rotation curves may be
affected by beam smearing, as in the case of NGC 5585 (compare with
the optical rotation curve by Blais-Ouellette et al. 1999). In the
case of DDO 168 the published rotation curve may be too low in the
inner parts and the model rotation curve is in fact consistent with
the observations. The fits for the other dwarf galaxies in Table~2 are
quite acceptable.}

\item[(2)]{The model rotation curve shows large wiggles that are not
present in the observed rotation curve. This effect is seen for
NGC~2841, NGC~2903 and NGC~300. In estimating the contribution of the
gas to the rotation velocity, we have used a radial profile that is an
azimuthal average of the HI distribution.  Features such as spiral
arms or blobs of HI, however, may cause bumps in the radial
profile. In the model rotation curve strong features will then show
up, because the scaling of HI amplifies the irregularities in the HI
distribution. Indeed the HI map of NGC 2903 shows
clear spiral structure and blobs related to the wiggles.
For NGC 2841 and NGC 300 the situation is less clear.
The presence of wiggles in the rotation curves shows that
the distribution of dark matter can not be an exact copy of the
distribution of HI. At best, the smeared-out distribution of HI is
a tracer of dark matter.}

\item[(3)]{The model rotation curve drops below the observed rotation
curve at large radii. Seven galaxies fall in this category (NGC~7331,
5033, 6674, 5371, 1560, 55, and also NGC~300). In Section~3 we
noted that for large galaxies HI flux may be missed in the 21-cm
synthesis observations. Obviously this could be a contributing factor
to a decline of the HI rotation curve in the outer parts. In the same
vein, hydrogen may be partly ionised in the outer regions due to
photons of intergalactic origin. This possibility has been discussed
for NGC 3198 by Maloney (1993). Irrespective of the presence of
molecular gas, the observed HI surface density may therefore not
always trace the total gas density in the outer parts.}

\end{enumerate}

We conclude from the comparisons in Fig.~\ref{rotcur} and the results
above that the rotation curves of two-thirds of the galaxies in our
sample, spirals and dwarfs alike, can be fitted with a mass model
based on the distribution of light and the distribution of HI,
provided that the scale factors ${\rm M}_\odot / {\rm L_{B\odot}}$ and
$\Sigma_{\rm dark}/\Sigma_{\rm HI}$ can be chosen freely. The values
we find for the mass-to-light ratios of the stellar component(s) agree
well with the values given in Broeils (1992a) and Begeman (1987). From
Fig.~\ref{rotcur} one sees that many fits are nearly maximum disc fits
(van Albada \& Sancisi, 1986). This is indeed a remarkable property of
this type of modeling: scaling of HI to represent the dark component
only works in combination with maximal discs. We have verified this by
gradually lowering the disc mass. Decreasing the disc mass by more
than $\sim 10\%$ below its maximum value results in a substantially
smaller core radius of the dark component, and such a dark component
can then no longer be represented by a scaled-up HI distribution.

\begin{figure}
\begin{center}
\leavevmode
\hbox{%
\epsfxsize=7cm
\epsffile{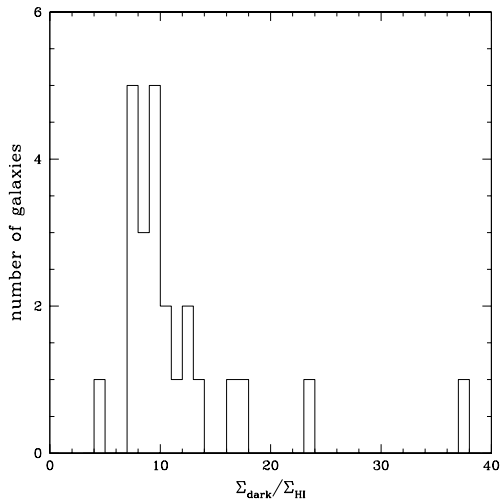}}
\end{center}
\caption{Frequency  distribution of the scaling factor 
$\Sigma_{\rm dark}/\Sigma_{\rm HI}$ found for
the galaxies in our sample.}\label{freq}
\end{figure}

In Fig.~\ref{freq} the frequency distribution of the scaling factor is
shown. It is strongly peaked around a value of approximately
9. Correcting for primordial helium we find that the dark matter
surface density is approximately 6.5 times the gas surface density
(i.e. HI + He). As can be seen from Table~2, the HI scaling factor
does not change much with galaxy type, although the scatter increases
towards earlier types.

Two galaxies, NGC~2841 (23) and NGC~2903 (37), have $\Sigma_{\rm
dark}/\Sigma_{\rm HI}>20$. These galaxies are the two most
compact ones in the sense that they have small scale lengths in
comparison to their maximum rotation velocities; they therefore
require a relatively massive dark halo. Their HI surface densities do
not differ much from those of other galaxies and this results in an HI
scale factor that is above average.

\section{Discussion}

As shown above, in general the distribution of HI mimics the
distribution of dark matter remarkably well, provided the contribution
of the disc to the rotation curve is close to maximal. We now examine
more closely which property, or properties, of the HI layer in
galaxies play a role in this.

\subsection{Bias against small HI scale length}

The rotation curves of the galaxies in our sample extend to well
beyond the radius of 2.2 scale lengths where the rotation curve of the
luminous component reaches its maximum. For a good match, we therefore
require a rotation curve for the HI component rising to well beyond
$2.2 h_{\rm stars}$, in order to compensate for the decline of the
rotation curve of the stars. Put differently, the radial `scale
length' of the HI distribution must be large, about 0.4 times the
outer radius of the HI disc, $R_{\rm out}$, or even larger. For the
galaxies in our sample we find an average $h_{\rm HI} \simeq 0.33
R_{\rm out}$, only barely enough for HI scaling to work.  Figure~1
shows that for most galaxies the HI surface density profiles curve
downwards with increasing radius. This is why for several galaxies
scaling of HI gives a poor fit in the outer parts, with the model
curve falling below the observed rotation curve (see e.g., NGC~7331
and 5371). Apparently, with the present 21-cm line sensitivity we may
be reaching the border of the region where scaling of atomic hydrogen
to fit the rotation curve works.  Extension of the rapidly declining
HI surface density profiles to larger radii would result in
progressively poorer fits.

It is worth noting that our selection in favour of a large radial
extent in HI excludes galaxies with small HI scale lengths, for the
following reason. The HI surface density in the inner region of spiral
galaxies, leaving aside a possible central depression, varies only by
a modest amount from galaxy to galaxy (Broeils 1992a, Rhee 1996 and
Fig.~\ref{profHI}). In the mean for our sample $\Sigma_{HI}(0) \simeq
6 {\rm M_{\odot}/pc^2}$.  The smallest HI surface densities measured
in the outer regions are set by the sensitivity of the observations.
For our sample typically: $ \Sigma_{\rm HI}(R_{\rm out}) \simeq 0.3
{\rm M_{\odot}}/{\rm pc}^2$. The corresponding range in $\Sigma_{\rm
HI}$ is therefore about a factor of 20 for most galaxies in our sample,
corresponding to three scale lengths for an exponential distribution.
As a result our sample is biased against galaxies with $R_{\rm out} /
h_{\rm HI}$ substantially larger than 3. Obviously, for such galaxies
scaling of HI to fit the rotation curve does not work.

\subsection{The HI scale factor}

Above we have shown that the success of our model fits is related
to rather specific properties of the HI distributions for the galaxies
in our sample. Could a selection effect also be responsible for the
small range in the scale factor seen in Figure~\ref{freq}?  Below we
shall argue that this is unlikely.

Let us assume that the mass distribution in a galaxy can be
represented by two exponential discs: one for the stars and one for
the dark matter.  The maximum rotation velocity of an exponential disc
is proportional to the product of the scale length and the central
surface density.  One observes a more or less flat rotation curve when
the maximum velocities of both discs are similar, i.e. when

\begin{equation}
\Sigma_{\rm stars}(0) h_{\rm stars} \approx \Sigma_{\rm dark} (0) 
h_{\rm dark},
\end{equation}
 
\noindent where $\Sigma_{\rm stars}(0)$ and $\Sigma_{\rm dark}(0)$ are
the central surface densities of respectively the stellar disc and the
dark halo, and where $h_{\rm stars}$ and $h_{\rm dark}$ indicate their
scalelengths.  This equation is a description of the disc-halo
conspiracy (van Albada \& Sancisi, 1986).

If we assume that the surface density profile of the dark matter is a
scaled version of the HI surface density, we can write for the scale
factor $f$:

\begin{equation}
f =\frac{\Sigma_{\rm dark}(R)}{\Sigma_{\rm HI}(R) }
\approx \frac{\Sigma_{\rm stars}(0)}{\Sigma_{\rm HI}(0)}
\frac{h_{\rm stars}}{h_{\rm dark}},
\end{equation}

\noindent where $\Sigma_{HI}(0)$ is a measure of the maximum HI
surface density, ignoring a possible central depression.  The small
spread in the scale factor $f$ therefore implies a coupling of the
scale lengths of stars and dark matter, as well as a coupling of the
surface mass densities of stars and HI.

As mentioned above, the maximum surface density of the HI disc of high
surface brightness galaxies does not vary much from galaxy to
galaxy. Similarly, for luminous spiral galaxies the central surface
brightness of the disc is remarkably constant (Freeman,
1970). Assuming some constant value for the mass-to-light ratio, this
would result in a fairly small variation in the central surface density
of the stellar disc.  Thus, the spread in the first factor in
Equation~(5) may well be small. This suggests that if the ratio of
scale lengths of the dark disc and the stellar disc were approximately
constant, then the scale factor $f$ would be more or less constant.
We note that, when calculating total surface densities from observed
rotation curves, we indeed found a more or less exponential surface
density profile in the outer parts. The scale length of this profile
was in many cases about 4 to 5 times the optical scale length. Thus
for the HSB galaxies in our sample, there appears to be a coupling
between the linear scales of the disc and the dark halo. Similar
results were found by Sackett (1997). Consequently, the expected
spread in $f$ is small for HSB galaxies.

Our sample also includes several dwarf galaxies, with central surface
brightnesses lower than those of the more luminous galaxies. Their HI
surface densities are comparable to the larger galaxies
however. Although we do not know how the stellar $M/L$ values of
dwarfs and more luminous galaxies differ, it seems unlikely that
equation~(5) would apply. The similarity of the scale factor $f$ for
dwarfs to that of the more luminous galaxies therefore seems a
significant result that can not easily be explained in terms of a
selection effect in our sample.

Swaters (1999) has applied HI scaling to a sample of 35 late-type
dwarf galaxies. With one or two exceptions he finds excellent
fits. Here too discrepancies, if any, occur mainly in the outermost
regions. The required scaling factors for HI show a considerable
spread, with a peak between 3 and 6. In the mean they are about a
factor of two smaller than the scaling factors found above.

\section{Summary \& Conclusions}

We have fitted mass models to the observed rotation curves of 24
galaxies under the assumption that the dark matter surface density is
proportional to the HI surface density. For about two thirds of the
galaxies we obtain good fits to the data. After correction for
primordial Helium the required scaling factor is about 7. The main
discrepancies occur at large radii, with the fitted rotation curves
falling below the observed ones. This results from a fairly rapid
decline of the HI surface density in the outermost regions.

For the HI data used here, the radial density profiles of HI typically
extend to about 3 scale lengths, i.e. close to or just beyond the
radius where the rotation curve of the HI component is expected to
decline. The good fits are therefore somewhat coincidental.  More
sensitive HI observations would reach the declining part of the
rotation curve of the HI component for several galaxies in our sample
(provided the HI surface density remains exponential further out). If
the observed circular velocity of the galaxy remains constant with
radius it would then no longer be possible to fit such a flat rotation
curve with a scaled up distribution of HI. Furthermore, in several
galaxies scaling up of HI leads to rotation curves with wiggles, and
in some galaxies not included in this paper the neutral hydrogen
density drops rather abruptly around the optical radius.

Therefore, in many cases, there seems to be little or no relation
between HI and dark matter.  Note, however, that for a number of
reasons (ionisation by the intergalactic radiation field, presence of
molecular gas and missing HI flux) the observed HI densities may be
lower than the actual gas densities.  Because of these complications
and the selection effects sketched above, it is not possible to
conclude here that there is a real coupling between HI and dark matter
in spiral galaxies.

\section*{Acknowledgments}

We thank Kor Begeman, Adrick Broeils and Floris Sicking for kindly
providing their data.

\end{document}